\documentclass[conference]{IEEEtran}
\IEEEoverridecommandlockouts

\usepackage{cite}
\usepackage{amsmath,amssymb,amsfonts}
\usepackage{algorithmic}
\usepackage{graphicx}
\usepackage{textcomp}
\usepackage{xcolor}
\def\BibTeX{{\rm B\kern-.05em{\sc i\kern-.025em b}\kern-.08em
    T\kern-.1667em\lower.7ex\hbox{E}\kern-.125emX}}
\usepackage{algorithm}
\usepackage[export]{adjustbox}
\usepackage{tikz}
\usepackage{array}
\usepackage{textcomp}
\usepackage{multirow}
\usepackage{tabularx,booktabs} 
\usepackage{verbatim}
\usepackage{cite}
\usepackage{makecell}
\usepackage{url}
\usepackage{relsize} 
\usepackage{color}
\usepackage{graphicx}
\usepackage{float}
\usepackage{enumitem}
\usepackage{bbm}
\usepackage{bm}
\usepackage{stfloats}
\usepackage{balance}
\usepackage[justification=centering]{caption}
\makeatletter
\renewcommand{\maketag@@@}[1]{\hbox{\m@th\normalsize\normalfont#1}}%
\makeatother
\usepackage{hyperref}

\begin{document}

\title{Hetero-Net: An Energy-Efficient Resource Allocation and 3D Placement in Heterogeneous LoRa Networks via Multi-Agent Optimization}

\author{
  Abdullahi Isa Ahmed\textsuperscript{1}, Ana-Maria Dr\u{a}gulinescu\textsuperscript{2},
  El Mehdi Amhoud\textsuperscript{1} \\
  \textsuperscript{1}College of Computing, Mohammed VI Polytechnic University (UM6P), Benguerir, Morocco. \\
  \textsuperscript{2}Telecommunications Department, Politehnica Bucharest, Bucharest, Romania. \\
  Emails: {\{abdullahi.isaahmed, elmehdi.amhoud\}@um6p.ma, ana.dragulinescu@upb.ro}
}

\maketitle

\begin{tikzpicture}[remember picture,overlay]
\node[anchor=north,yshift=-20pt, text=black] at (current page.north) {\parbox{\dimexpr\textwidth-\fboxsep-\fboxrule\relax}{
\centering\footnotesize 
This paper has been accepted for publication in the 2026 IEEE International Conference on Communications (ICC) Workshop \textcopyright 2026 IEEE.\\
Please cite it as: A.I. Ahmed, A.M. Dr\u{a}gulinescu and E. M. Amhoud, “Hetero-Net: An Energy-Efficient Resource Allocation and 3D Placement in Heterogeneous LoRa Networks via Multi-Agent Optimization,” in IEEE International Conference on Communications (ICC) Workshop, May 2026.}};
\end{tikzpicture}

\begin{abstract}
The evolution of Internet of Things (IoT) into multi-layered environments has positioned Low-Power Wide Area Networks (LPWANs), particularly Long Range (LoRa), as the backbone for connectivity across both surface and subterranean landscapes. However, existing LoRa-based network designs often treat ground-based wireless sensor networks (WSNs) and wireless underground sensor networks (WUSNs) as separate systems, resulting in inefficient and non-integrated connectivity across diverse environments. To address this, we propose Hetero-Net, a unified heterogeneous LoRa framework that integrates diverse LoRa end devices with multiple unmanned aerial vehicle (UAV)-mounted LoRa gateways. Our objective is to maximize system energy efficiency through the joint optimization of the spreading factor, transmission power, and three-dimensional (3D) placement of the UAVs. To manage the dynamic and partially observable nature of this system, we model the problem as a partially observable stochastic game (POSG) and address it using a multi-agent proximal policy optimization (MAPPO) framework. An ablation study shows that our proposed MAPPO Hetero-Net significantly outperforms traditional, isolated network designs, achieving energy efficiency improvements of 55.81\% and 198.49\% over isolated WSN-only and WUSN-only deployments, respectively.
\end{abstract}

\begin{IEEEkeywords}
Energy-efficient, multi-agent, UAV placement, resource allocation, heterogeneous LoRa.
\end{IEEEkeywords}

\section{Introduction} \label{section1}
In recent years, low-power wide area networks (LPWANs) have emerged as a key enabling technology for autonomous Internet of Things (IoT) applications, offering extended range and energy efficiency for a wide array of devices \cite{jouhari2023survey}. Among these technologies, Long Range (LoRa) has proven particularly successful due to its low power consumption, low data rate, and long-range capabilities, making it ideal for large-scale, distributed sensor deployments. 

Traditionally, LoRa networks have been designed for ground-based wireless sensor networks (WSNs). However, emerging demands from critical sectors, such as smart agriculture, mining, and infrastructure monitoring, are driving the need for connectivity in more challenging environments. In particular, wireless underground sensor networks (WUSNs) have become essential for applications like real-time pipeline monitoring. In these scenarios, energy-efficient and reliable communication is vital despite the severe signal attenuation and path loss caused by underground propagation \cite{10297305}. These evolving requirements highlight the growing importance of supporting both surface and subterranean connectivity within a unified framework.

Extensive literature exists on optimizing LoRa networks for ground-based WSNs, with a strong focus on improving energy efficiency \cite{jouhari2023deep, yu2020multi, 10978199}. For instance, the authors of \cite{yu2020multi} proposed a multi-agent Q-learning algorithm to jointly optimize transmission power and spreading factor in uplink LoRa communication. Similarly, the authors in \cite{10978199} introduced a distributed reinforcement learning-based scheme for transmission power and channel selection. Conversely, research on WUSNs has addressed the unique constraints of underground environments. In \cite{zhang2025parameter}, an ``Energy Saver'' configuration method was proposed, which uses real-time soil moisture data to estimate path loss and adapt LoRa parameters for optimal energy usage. More recently, a multi-agent reinforcement learning (MARL) approach was adopted in \cite{10297305} to allocate spreading factors in direct-to-satellite LoRa scenarios, aiming to reduce co-spreading factor interference in dense WUSN deployments.

Although these studies provide valuable insights within their respective domains, they fall short of addressing heterogeneous LoRa deployments that span both ground and underground environments. Moreover, conventional ground-based gateway deployments suffer from limited coverage and non-line-of-sight (NLoS) challenges, whereas satellite-based solutions, such as those in \cite{10297305}, often introduce high latency, complex infrastructure requirements, and increased power consumption. These limitations highlight the need for a flexible, integrated approach capable of adapting to dynamic network conditions and diverse environmental constraints.

In this paper, we propose Hetero-Net, a heterogeneous LoRa-based network architecture that utilizes unmanned aerial vehicles (UAVs) as mobile gateways to collect data from both WSN and WUSN LoRa end devices. Our objective is to maximize system energy efficiency by jointly optimizing the three-dimensional (3D) placement of UAV-mounted gateways, spreading factor, and transmission power across all LoRa end devices. To solve this complex coordination task, we model the optimization problem as a partially observable stochastic game (POSG) and leverage multi-agent proximal policy optimization (MAPPO) algorithm. To the best of our knowledge, this is the first study to address energy efficiency in heterogeneous LoRa deployments involving both ground and underground sensor networks in a multi-agent setting. Our main contributions are summarized as follows:
\begin{itemize}
\item We propose Hetero-Net, a unified LoRa-based framework that supports both ground-based WSNs and WUSNs. This framework accounts for the physical disparities between these layers by incorporating realistic device positioning and channel modeling for ground-to-air (G2A) and underground-to-air (UG2A) communication links.
\item We formulate the joint problem of UAV 3D placement and LoRa parameter allocation as a single energy efficiency optimization task, modeled as a POSG. Unlike existing literature that optimizes these parameters in isolation, our framework incorporates tailored state representations, action spaces, and reward functions to balance the divergent requirements of surface and subterranean communication.
\item We demonstrate through extensive simulations that the proposed MAPPO-based approach outperforms existing deep reinforcement learning (DRL) and non-DRL baselines in terms of energy efficiency. Furthermore, ablation studies confirm that jointly optimizing this heterogeneous LoRa deployment significantly improves overall system performance.
\end{itemize}
The remainder of the paper is structured as follows. Section~\ref{section2} presents the system model and the formal problem formulation. Section~\ref{section3} details the proposed MAPPO-based solution approach. Simulation results and performance analysis are provided in Section~\ref{section4}. Finally, Section~\ref{section5} concludes the paper and outlines directions for future research.
\vspace{-2.5pt}

\section{System Model} \label{section2}
\begin{figure}
  \centering 
\includegraphics[width=1.0\linewidth]{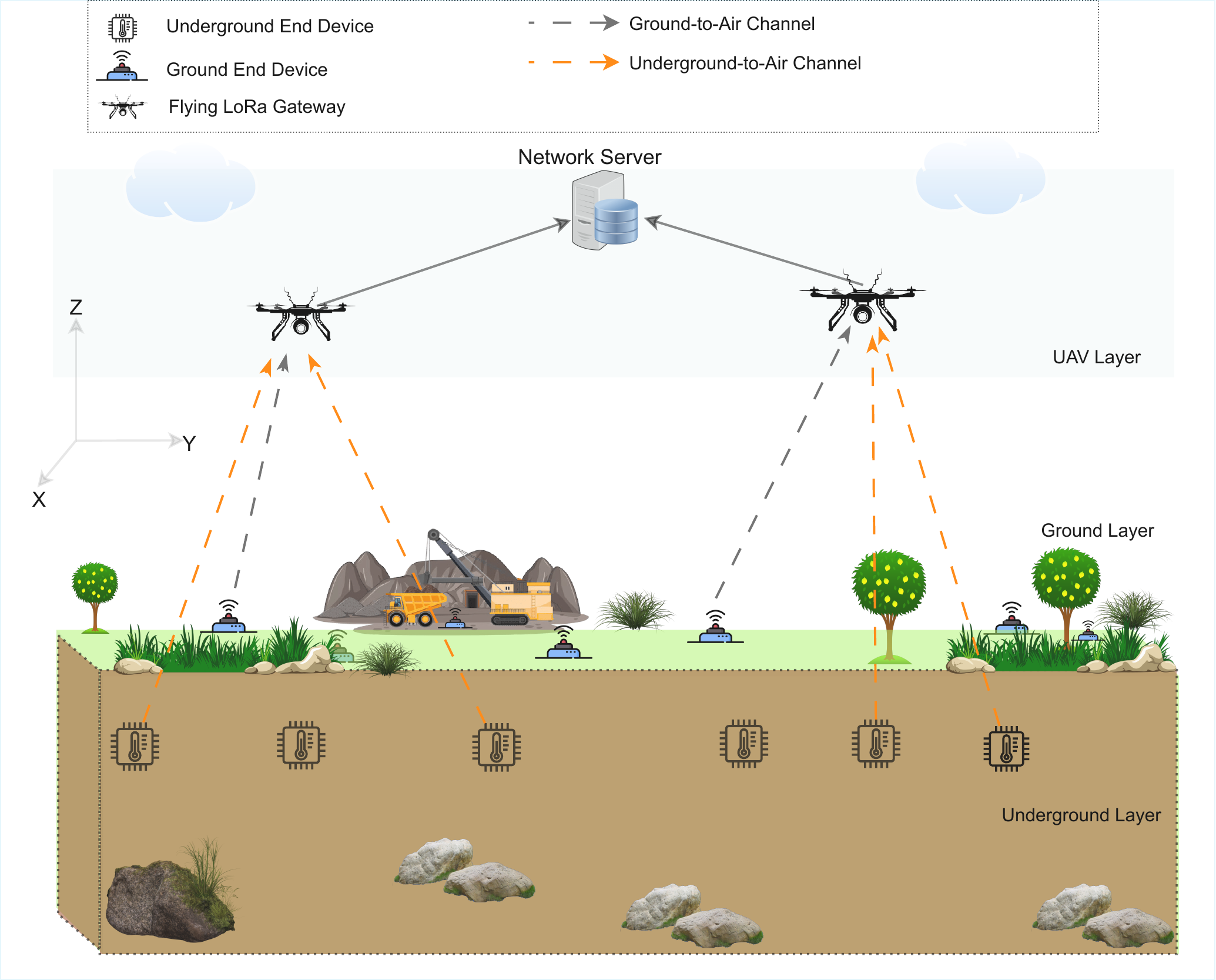}
  \caption{The studied system model.}
  \label{fig_system_model}
  \vspace{-12pt}
\end{figure}

In this work, we consider a LoRa uplink heterogeneous communication system comprising multiple ground and underground LoRa end devices, as well as multiple UAVs. Specifically, the system includes $U$ non-terrestrial UAVs equipped with LoRa gateways, denoted by the set $\mathcal{U} \triangleq \{1, 2, \ldots, U\}$, which serve as aerial data collection and forwarding stations. Multiple LoRa end devices are distributed across two terrestrial layers, denoted by the set $\mathcal{V} \triangleq \{1, 2, \ldots, V\}$. Each end device $v \in \mathcal{V}$ is deployed in one of two layers, i.e., $k=0$ for the underground layer or $k=1$ for the ground layer. The end devices are geographically organized into $C$ clusters, represented by the set $\mathcal{C} \triangleq \{1, 2, \ldots, C\}$, where each cluster corresponds to a point of interest (POI) in the deployment area. We assume a one-to-one mapping between UAVs and clusters, such that UAV $u \in \mathcal{U}$ is assigned to serve cluster $c = u$. We denote $\mathcal{V}_{u}$ as the set of all end devices associated with UAV $u$, which is known a priori based on their proximity to the corresponding POI. Furthermore, $\mathcal{V}^{k}_{u} \subseteq \mathcal{V}_{u}$ represents the subset of end devices served by UAV $u$ that are located at layer $k \in \{0,1\}$. Clearly, $\mathcal{V}_{u} = \mathcal{V}^{0}_{u} \cup \mathcal{V}^{1}_{u}$ and $\mathcal{V}^{0}_{u} \cap \mathcal{V}^{1}_{u} = \emptyset$. An illustration of the studied system model is provided in Fig.~\ref{fig_system_model}.

Moreover, the static wireless ground and underground LoRa end devices are distributed within a target area $\mathcal{A}_c$. Within each cluster $u$, the horizontal coordinates of all associated end devices at layer $k$, denoted by $\mathcal{G}_u^{k} = \{(x_v^{k}, y_v^{k}) : v \in \mathcal{V}_u^{k}\}$, follow a bi-variate normal spatial distribution centered at the cluster centroid $\boldsymbol{\mu}_u$ with variance $\sigma_u^2$, i.e., $\mathcal{G}_u^{k} \sim \mathcal{N}(\boldsymbol{\mu}_u, \sigma_u^2 \mathbf{I}_2)$. Consequently, each end device $v \in \mathcal{V}_u^{k}$ is located at 3D position $\mathbf{g}_v^{k} =[x_v^{k}, y_v^{k}, h_v^{k}]$, where the altitude component is $h_v^{0} = -d_b$, where $d_b$ denote the burial depth for underground devices, and $h_v^{1} = 0$ for ground-level devices. Furthermore, each UAV gateway $u \in \mathcal{U}$ operates at an altitude $h_u > 0$, with a 3D position defined as $\mathbf{q}_u =[x_u, y_u, h_u]$.

\subsection{Channel Model}
\subsubsection{Underground-to-Air (UG2A) Model} 
Following the model in \cite{9295338}, the total UG2A path loss from an underground LoRa end device $v \in \mathcal{V}^{0}_{u}$ to UAV gateway $u$ consists of three primary components: 1) the above-ground air attenuation $L^{\mathrm{air}}_{u,v}$, 2) the underground soil attenuation $L^{\mathrm{soil}}$, and 3) the refraction loss at the air-soil interface $L^{\mathrm{ref}}$. First, the underground soil attenuation $L^{\mathrm{soil}}$ model is expressed as \cite{10297305}
\vspace{-5pt}
\begin{equation}
   L^{\mathrm{soil}} = \left( \frac{2\beta d_{p}}{e^{-\alpha d_{p}}} \right)^2,  
   \label{equation_soil}
\end{equation}
where the attenuation constant $\alpha$ and phase shifting constant $\beta$ are given by
\vspace{-5pt}
\begin{align}
    \alpha &= 2\pi f \sqrt{\frac{\mu_r \mu_0 \epsilon' \epsilon_0}{2} \left[ \sqrt{1 + \left( \frac{\epsilon''}{\epsilon'} \right)^2} - 1 \right]}, \label{equation_alpha}  \\
    \beta &= 2\pi f \sqrt{\frac{\mu_r \mu_0 \epsilon' \epsilon_0}{2} \left[ \sqrt{1 + \left( \frac{\epsilon''}{\epsilon'} \right)^2} + 1 \right]}. \label{equation_beta}
\end{align}

Here, $f$ is the operating frequency, $\mu_r$ is the relative permeability of the soil, $\mu_0$ is the permeability of free space, and $\epsilon_0$ is the permittivity of free space. The parameters $\epsilon'$ and $\epsilon''$ are the real and imaginary parts of the soil's complex relative permittivity,  respectively. Note that $\epsilon_r = \epsilon' - j\epsilon''$, where the real part $\epsilon'$ represents energy storage capability, while $\epsilon''$ quantifies energy dissipation due to losses in the soil. The parameter $d_{p} = d_b / \cos\left(\arcsin\left(1/\sqrt{\epsilon'}\right)\right)$ in Eq.\eqref{equation_soil} represents the actual underground path length, where $d_b$ is the burial depth assuming near-normal propagation. The specific values of these parameters are detailed in \cite{4895263}. Next, to account for the above-ground air attenuation, we model $L^{\mathrm{air}}_{u,v}$ for underground device $v \in \mathcal{V}^{0}_{u}$ as
\vspace{-5pt}
\begin{equation}
    L^{\mathrm{air}}_{u,v} = \left( \frac{4\pi f}{c_0} \right)^{2} d_{u,v}^{\eta^0},
    \label{equation_air}
\end{equation}
where $\eta^0$ is the path loss exponent, $c_0$ is the speed of light, and $d_{u,v}$ denotes the Euclidean distance between underground device $v$ at position $\mathbf{g}^{0}_{v}$ and UAV $u$ at position $\mathbf{q}_u$. Lastly, following the approach in \cite{10297305}, we neglect the refraction loss at the air-soil interface, setting $L^{\mathrm{ref}}_{u,v} = 1$. This simplification is justified because most electromagnetic energy is successfully refracted when waves propagate from a high-density medium (soil) to a lower-density medium (air), making the refraction loss negligible. Consequently, we model the overall UG2A path loss as
\begin{equation}
L^{\mathrm{UG2A}}_{u,v} = L^{\mathrm{soil}} \cdot L^{\mathrm{air}}_{u,v}.
\label{equation_total_pathloss}
\end{equation}

\subsubsection{Ground-to-Air (G2A) Model} 
The G2A communication for ground LoRa end devices $v \in \mathcal{V}^{1}_{u}$ follows a probabilistic path loss model that accounts for both Line-of-Sight (LoS) and Non-Line-of-Sight (NLoS) conditions. The occurrence of LoS or NLoS depends on the environmental characteristics, the elevation angle, and the presence of obstacles. According to \cite{8765827}, the average path loss of the G2A channel between ground end device $v$ and UAV $u$ is modeled as
\vspace{-5pt}
\begin{equation}
\begin{split}
     L^{\mathrm{G2A}}_{u,v} = 20\log \left(\frac{4\pi f d^{\mathrm{G2A}}_{u,v}}{c_0} \right) +  P^{\mathrm{LoS}}_{u,v} \eta^\mathrm{G2A}_{\mathrm{LoS}} \\
     + \left(1 - P^{\mathrm{LoS}}_{u,v} \right) \eta^\mathrm{G2A}_{\mathrm{NLoS}},
\end{split}
\label{equation_g2a_pathloss}
\end{equation}
where $d^{\mathrm{G2A}}_{u,v} = \|\mathbf{q}_u - \mathbf{g}^{1}_{v}\|$ is the Euclidean distance between UAV $u$ and ground end device $v$, while $\eta^\mathrm{G2A}_{\mathrm{LoS}}$ and $\eta^\mathrm{G2A}_{\mathrm{NLoS}}$ are environment-dependent parameters that representing the average additive losses for LoS and NLoS links, respectively. Additionally, $P^{\mathrm{LoS}}_{u,v}$ is the LoS probability between ground end device $v$ and UAV $u$, which can be expressed as
\vspace{-5pt}
\begin{equation}
    P^{\mathrm{LoS}}_{u,v} = \frac{1}{1 + \phi \exp\left(-\varphi \left[\theta^{\mathrm{G2A}}_{u,v} - \phi \right]\right)},
    \label{equation_los_prob}
\end{equation}
where $\phi$ and $\varphi$ are environment-dependent parameters that characterize the propagation environment. Specifically, larger $\varphi$ values indicate faster transitions from NLoS to LoS, typical of open or rural environments, while smaller $\varphi$ reflects smoother transitions in dense urban areas. Similarly, larger $\phi$ values correspond to scenarios requiring higher elevation angles to achieve LoS probability, as encountered in urban environments with tall buildings. The elevation angle $\theta^{\mathrm{G2A}}_{u,v}$ between ground end device $v$ and UAV $u$ is given by
\vspace{-5pt}
\begin{equation}
    \theta^{\mathrm{G2A}}_{u,v} = \arcsin\left(\frac{h_u}{\|\mathbf{q}_u - \mathbf{g}^{1}_{v}\|}\right).
    \label{eq:elevation_angle}
\end{equation}

Finally, we model the channel gain for both communication links. The UG2A channel gain between UAV $u$ and underground device $v \in \mathcal{V}^{0}_{u}$ is given by $g^{0}_{u,v} = 10^{-L^{\mathrm{UG2A}}_{u,v}/10}$, while the G2A channel gain between UAV $u$ and ground device $v \in \mathcal{V}^{1}_{u}$ is expressed as $g^{1}_{u,v} = 10^{-L^{\mathrm{G2A}}_{u,v}/10}$. Consequently, the channel gain for any device $v \in \mathcal{V}^{k}_{u}$ where $k \in \{0,1\}$ can be unified as
\begin{equation}
    g^{k}_{u,v} = \begin{cases}
        10^{-L^{\mathrm{UG2A}}_{u,v}/10} & \text{if } k=0 \text{ (underground)}, \\
        10^{-L^{\mathrm{G2A}}_{u,v}/10} & \text{if } k=1 \text{ (ground)}.
    \end{cases}
    \label{eq:unified_channel_gain}
\end{equation}

\subsection{Power Consumption Model}
\subsubsection{LoRa End Device Power Consumption Model}
The power consumption of a LoRa-enabled end device $v \in \mathcal{V}^{k}_{u}$ at layer $k$ consists mainly of two components: the transmission power $P^{\text{tx}}_{v}$ and the circuit power $P^{\text{circuit}}_{v}$. Therefore, the total power consumption of end device $v$ during uplink transmission is modeled as \cite{jouhari2023deep}
\vspace{-5pt}
\begin{equation}
P_{v} = P^{\text{tx}}_{v} + P^{\text{circuit}}_{v}.
\label{eq:ed_power}
\end{equation}

\subsubsection{UAV Hovering Power Model}
In addition to the transmission power of LoRa end devices, a UAV also consumes power during various aerial operations such as hovering and steady flight. In our proposed scenario, we consider only the hovering power. The induced hover power for a multirotor UAV $u \in \mathcal{U}$ can be expressed as \cite{10175052}
\vspace{-10pt}
\begin{equation}
    P_{u,\text{hover}} = (1 + k_{\text{ind}}) \cdot W_u \cdot \sqrt{\frac{W_u}{2 \rho \bar{j} A_{\text{rotor}}}},
\label{eq:uav_hover_power}
\end{equation}
where $W_u$ is the weight of UAV $u$ in Newtons, $k_{\text{ind}}$ is the incremental induced-power factor, $\rho$ is the air density in kg/m³, $\bar{j}$ is the number of rotors, and $A_{\text{rotor}}$ is the rotor disk area per rotor in m². The factor $(1 + k_{\text{ind}})$ accounts for non-ideal effects, such as rotor wake interactions and tip losses, that increase the required power beyond the theoretical minimum.

\subsection{Energy Efficiency Model}
To link the physical-layer transmission quality to the overall system performance, we define the energy efficiency of the LoRa UAV-assisted heterogeneous network by incorporating the signal-to-interference-plus-noise ratio (SINR), the achievable data rate, and the total system power consumption. As a first step, we characterize the uplink transmission quality between an end device and its associated UAV. Specifically, we consider an end device $v \in \mathcal{V}^{k}_{u}$ at layer $k$ served by UAV $u \in \mathcal{U}$. Therefore, the signal-to-noise ratio (SNR) between $v$ and $u$ is given by $\rho^{k}_{u,v} = \frac{P_{v} g^{k}_{u,v}}{\sigma^{2}}$, where $g^{k}_{u,v}$ denotes the channel gain from $v$ to $u$, $P_{v}$ is the transmit power of $v$, and $\sigma^{2}$ is the noise power.

The achievable uplink data rate from $v$ to $u$ follows the Shannon-Hartley model \cite{aczel1974shannon} and is given by
\begin{equation}
\label{equation_datarate}
\Re^{k}_{u,v} = BW_{v}\, \log_2 \!\left( 1 + \gamma^{k}_{u,v,n} \right),
\end{equation}
where $BW_{v}$ is the bandwidth allocated to end device $v$, $n \in \{7, 8, 9, 10, 11, 12\}$ is the spreading factor assigned to device $v$, and the SINR is
\vspace{-5pt}
\begin{equation}
\label{equation_sinr}
\gamma^{k}_{u,v,n}
= \frac{P_{v} g^{k}_{u,v}}{\displaystyle
\sum_{k' \in \{0,1\}}
\sum_{v' \in \mathcal{V}^{k'}_{u} \setminus \{v\}}
\delta_{v',v} \; P_{v'}\, g^{k'}_{u,v'} + \sigma^{2}}.
\end{equation}
Here, $\mathcal{V}^{k'}_{u}$ is the set of end devices served by UAV $u$ at layer $k' \in \{0,1\}$, and $\delta_{v',v}$ is an indicator variable that equals 1 when device $v'$ uses the same spreading factor as device $v$, and 0 otherwise. We assume perfect orthogonality between different spreading factors, such that only co-SF devices contribute to interference. Additionally, we neglect inter-UAV interference due to sufficient spatial separation between UAV clusters.

Finally, the overall energy efficiency of the system is defined as the sum of per-UAV energy efficiencies, representing the ratio of total throughput to total power consumption, and can be presented as
\vspace{-5pt}
\begin{equation}
      \eta_{\mathrm{EE}} = \sum_{u \in \mathcal{U}} \left[ \; \frac{\sum_{k \in \{0,1\}}  \sum_{v \in \mathcal{V}^{k}_{u}} \Re^{k}_{u,v}}{ \left(\sum_{k \in \{0,1\}}  \sum_{v \in \mathcal{V}^{k}_{u}} P_{v}\right) + P_{u,\text{hover}}} \; \right].
      \label{equation_ee}
\end{equation}
Our goal is to maximize the energy efficiency of the proposed heterogeneous LoRa network. To this end, we jointly optimize the spreading factor assignment, transmission power allocation, and 3D placement of the UAV-mounted gateways. Let the discrete sets of spreading factors and power levels be denoted by $\mathbf{\Phi}$ and $\mathbf{P}$, respectively. We define the UAV placement decision as $\mathbf{Q} \triangleq \{(x_u, y_u, h_u)\}_{u\in\mathcal{U}}$, the spreading factor assignment as $\mathbf{n} \triangleq \{n_v\}_{v\in\mathcal{V}}$, and the power allocation as $\mathbf{P}_{\text{tx}} \triangleq \{P_v\}_{v\in\mathcal{V}}$. Our constrained optimization problem is formulated as follows
\vspace{-7pt}
\begin{subequations}\label{formulated_problem}
\begin{align}
& \max_{\mathbf{n}, \mathbf{P}_{\text{tx}}, \mathbf{Q}} \quad \eta_{\mathrm{EE}} \label{constr:a} \\
\text{s.t.} \quad
& 0 \le x_u \le x^{\text{max}}, \quad \forall u \in \mathcal{U}, \label{constr:b} \\
& 0 \le y_u \le y^{\text{max}}, \quad \forall u \in \mathcal{U}, \label{constr:c} \\
& h_{\min} \le h_u \le h_{\max}, \quad \forall u \in \mathcal{U}, \label{constr:d} \\
& P_v \in \mathbf{P}, \quad \forall v \in \mathcal{V}, \label{constr:e} \\
& n_v \in \mathbf{\Phi}, \quad \forall v \in \mathcal{V}, \label{constr:f} \\
& \rho^{k}_{u,v} \ge \bar{\mho}_{n_v}, \forall v \in \mathcal{V}^{k}_{u}, \forall u \in \mathcal{U}, k \in \{0,1\}. \label{constr:g} 
\end{align}
\end{subequations}

Here, constraints~\eqref{constr:b} and~\eqref{constr:c} confine each UAV within the target area, while constraint~\eqref{constr:d} restricts the altitude to the operational range $[h_{\min}, h_{\max}]$. Constraints~\eqref{constr:e} and~\eqref{constr:f} ensure that transmission power and spreading factor selections come from the allowed discrete sets. Finally, constraint~\eqref{constr:g} guarantees feasibility by enforcing that the received SNR $\rho^{k}_{u,v}$ must exceed the minimum sensitivity threshold $\bar{\mho}_{n_v}$ corresponding to the assigned spreading factor for each device $v$. Note that the typical SNR thresholds for LoRa transceivers operating at $BW_v = 125$ kHz are summarized in Table~\ref{table:snr_threshold}.

The problem in~\eqref{formulated_problem} is a mixed-integer nonlinear program due to the discrete spreading factor and power selections, coupled with the non-convex nature of the wireless channel models. It is therefore known to be NP-hard in general. In the next section, we propose a MARL framework to obtain an efficient solution.

\section{Proposed Approach}\label{section3}
We tackle the optimization problem in \eqref{formulated_problem} using a MARL framework based on centralized training and decentralized execution (CTDE). Specifically, we adopt the MAPPO \cite{yu2022surprising} algorithm, an actor-critic method suitable for partially observable environments. Furthermore, we model the environment as a POSG, defined by the tuple: $\langle \mathcal{U}, \mathcal{S}, \{\mathcal{A}_u\}, \{\mathcal{O}_u\}, \mathcal{T}, \{\mathcal{Z}_u\}, \{\mathcal{R}_u\}, \gamma, \hat{\mu}_0 \rangle$, where $\mathcal{U}$ is the set of agents, $\mathcal{S}$ is the global state space, $\{\mathcal{A}_u\}$ and $\{\mathcal{O}_u\}$ are the individual action and observation spaces for each agent, $\mathcal{T}$ is the state transition function, $\{\mathcal{Z}_u\}$ are the observation functions, $\{\mathcal{R}_u\}$ are the individual reward functions, $\gamma$ is the discount factor, and $\hat{\mu}_0$ is the initial state distribution. 
\begin{table}[t]
    \centering
    {\small
    \caption{\small{SNR Thresholds $\bar{\mho}_{n}$ for Different Spreading Factors with $BW_v = 125$ kHz}}
    \label{table:snr_threshold}}
    \begin{tabular}{|c|c|c|c|c|c|c|} \hline
        Spreading Factor ($n$) & 7 & 8 & 9 & 10 & 11 & 12 \\ \hline
        SNR Threshold $\bar{\mho}_{n_v}$ (dB) & -7.5 & -10 & -12.5 & -15 & -17.5 & -20 \\
        \hline
    \end{tabular}
    \vspace{-15pt}
\end{table}

The proposed heterogeneous network-based MAPPO algorithm, referred to as MAPPO-Hetero-Net, involves the following elements:

\begin{itemize}
    \item \textbf{Agents ($\mathcal{U}$)}: The finite set of agents.
    \item \textbf{Actions ($\mathcal{A}_u$)}: Each agent $u \in \mathcal{U}$ has an individual action space $\mathcal{A}_u$. The action of agent $u$ is denoted as $a_u = [\Delta x_u, \Delta y_u, \Delta h_u, n, P] \in \mathcal{A}_u$. Here, $(\Delta x_u, \Delta y_u, \Delta h_u)$ are the normalized movement deltas in $[-1,1]$ scaled by the maximum step size $s_{\mathrm{step}}$. That is, $(\Delta x_u, \Delta y_u, \Delta h_u) = s_{\mathrm{step}} \cdot \tilde{\mathbf{d}}_u, \; \tilde{\mathbf{d}}_u \in [-1,1]^3$, while $n$ and $P$ are the spreading factor and power assignments, respectively. The joint action space is $\mathcal{A} = \prod_{u \in \mathcal{U}} \mathcal{A}_u$, with joint action $\mathbf{a} = (\mathbf{a}_1, \mathbf{a}_2, \ldots, \mathbf{a}_U)$. 
    
    \item \textbf{Observations ($\mathcal{O}_u$)}: Each agent $u$ receives a local observation $o_u \in \mathcal{O}_u$ based on its partial view of the environment. Therefore, the observation of agent $u$ is given by $o_u = [x_u, y_u, h_u, \{d_{u,v}, \rho^{k}_{u,v}, n_v, P_v\}_{v\in\mathcal{V}_{u}}]$, where $x_u, y_u, h_u$ represent the UAV's 3D position, $d_{u,v}$ is the Euclidean distance between UAV $u$ and end device $v \in \mathcal{V}_{u}$, $\rho^{k}_{u,v}$ is the received SNR at layer $k$, $n_v$ is the spreading factor, and $P_v$ is the transmit power. The joint observation is denoted $o = (o_1, o_2, \ldots, o_U)$. Note that we normalized all distance, SNR, spreading factor, and power features using min-max scaling per cluster to $[0,1]$ to enhance the learning process.

    \item \textbf{Rewards ($\mathcal{R}_u$)}: Each agent $u$ has an individual reward function $\mathcal{R}_u: \mathcal{S} \times \mathcal{A} \rightarrow \mathbb{R}$ that balances system-wide performance with local cluster efficiency. We first define a local energy efficiency metric for agent $u$ that reflects how well it manages the end devices in its associated cluster given by $ \eta_{\mathrm{EE}}^{\mathrm{local}}(u) = \frac{\sum_{k \in \{0,1\}} \sum_{v \in \mathcal{V}^{k}_{u}} \Re^{k}_{u,v}}{\left(\sum_{k \in \{0,1\}} \sum_{v \in \mathcal{V}^{k}_{u}} P_{v}\right) + P_{u,\text{hover}}}$, where $P_{u,\text{hover}}$ is the altitude-dependent hovering power of UAV $u$. Next, we incorporate the global energy efficiency $\eta_{\mathrm{EE}}$ from Eq.~\eqref{equation_ee}, which captures the system-level trade-off between total throughput and total consumed power across all agents. The reward function for agent $u$ combines both terms and is expressed as
    \vspace{-7pt}
    \begin{equation}
    \mathcal{R}_u(s, \mathbf{a}) = \omega\,\eta_{\mathrm{EE}} + (1-\omega)\,\eta_{\mathrm{EE}}^{\mathrm{local}}(u),
    \label{eq:step_reward}
    \end{equation}
    where the scalar $\omega \in [0,1]$ is a tunable weight that balances the global objective against local cluster efficiency.
\end{itemize}

Each agent $u$ follows a decentralized policy $\pi_u: \mathcal{O}_u \times \mathcal{A}_u \rightarrow [0,1]$, which maps its local observation to a probability distribution over actions. Therefore, the objective of each agent is to maximize its expected cumulative discounted reward, given as
\vspace{-5pt}
\begin{equation}
J_u(\pi_u) = \mathbb{E}_{\tau_u \sim \pi_u} \left[ \sum_{t=0}^{T} \gamma^t \mathcal{R}_u(\mathbf{s}[t], \mathbf{a}[t]) \right],
\end{equation}
where the expectation is taken over the trajectories $\tau_u$ generated by the agent $u$. In the cooperative setting, agents aim to maximize the objective of the team $J(\boldsymbol{\pi}) = \sum_{u=1}^{U} J_u(\pi_u)$.

MAPPO-Hetero-Net operates under the CTDE paradigm, where each agent $u \in \mathcal{U}$ maintains a decentralized policy network $\pi_{\theta_u}(a_u|o_u)$ mapping local observations to actions, while a centralized critic $V_\phi(s)$ estimates state values using the global state $s$ during training. Agents compute advantages via generalized advantage estimation (GAE) and update policies simultaneously by maximizing a clipped PPO objective with probability ratio $r_u(\theta_u) = \pi_{\theta_u}(a_u|o_u) / \pi_{\theta_u^{\text{old}}}(a_u|o_u)$, where $\pi_{\theta_u^{\text{old}}}(a_u|o_u)$ represents the old policy from the previous iteration with respect to which the ratio is computed. The ratio is constrained within $[1-\epsilon_{clip}, 1+\epsilon_{clip}]$ for stability, where the clipping parameter $\epsilon_{clip}$ prevents destructively large policy updates in a single step, ensuring monotonic improvement and training stability \cite{yu2022surprising}. During deployment, each UAV operates autonomously using only its policy $\pi_{\theta_u}(a_u|o_u)$ and local observation $o_u$, eliminating inter-agent communication requirements. A detailed discussion of the MAPPO algorithm and its computational complexity analysis can be found in \cite{li2024collaborative}.
\begin{figure*}
\centering
\begin{tabular}{ccc}
    \vspace{-6pt}
    \includegraphics[width=0.325\textwidth]{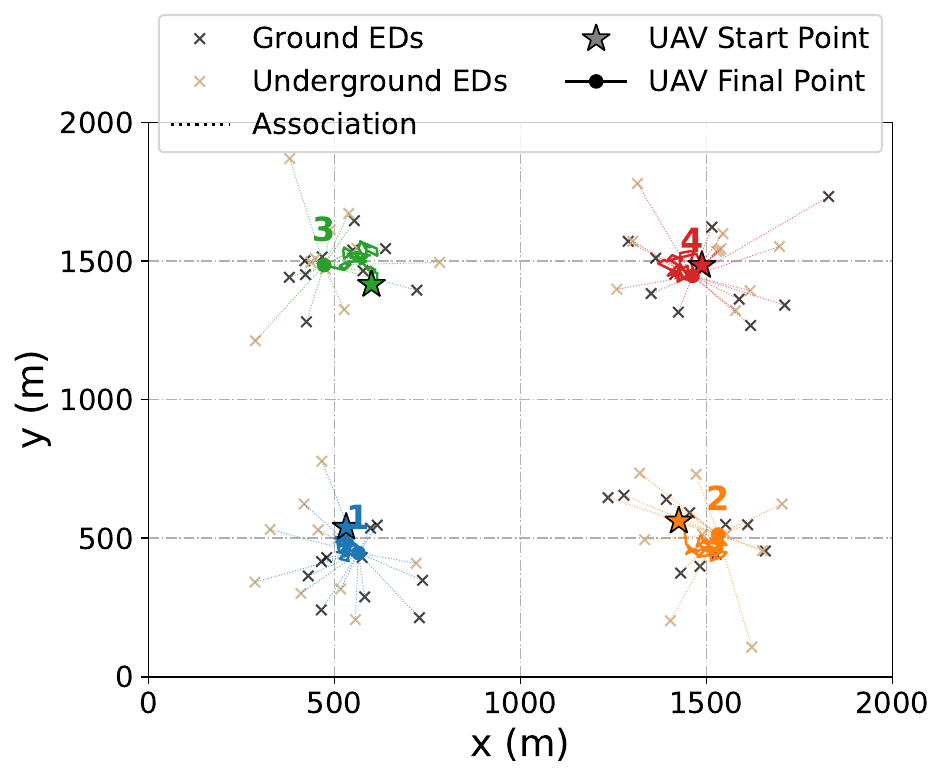} & 
    \includegraphics[width=0.325\textwidth]{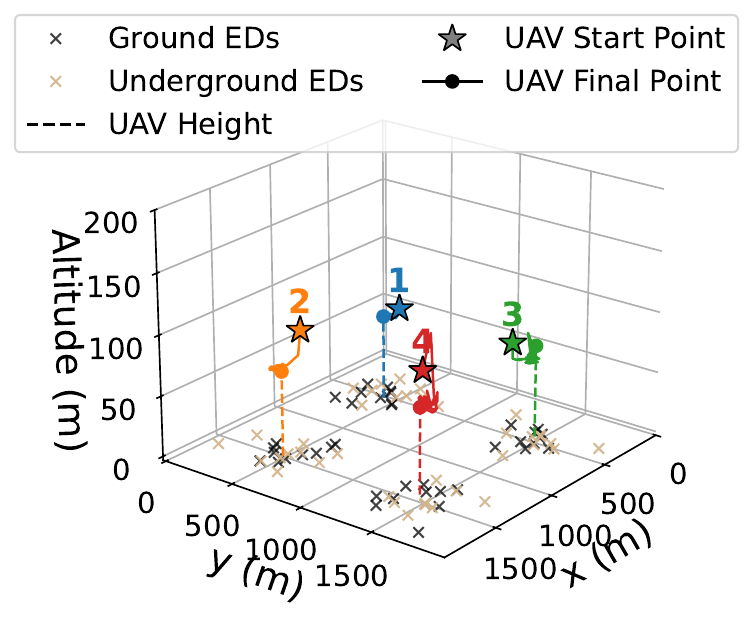} & 
    \includegraphics[width=0.325\textwidth]{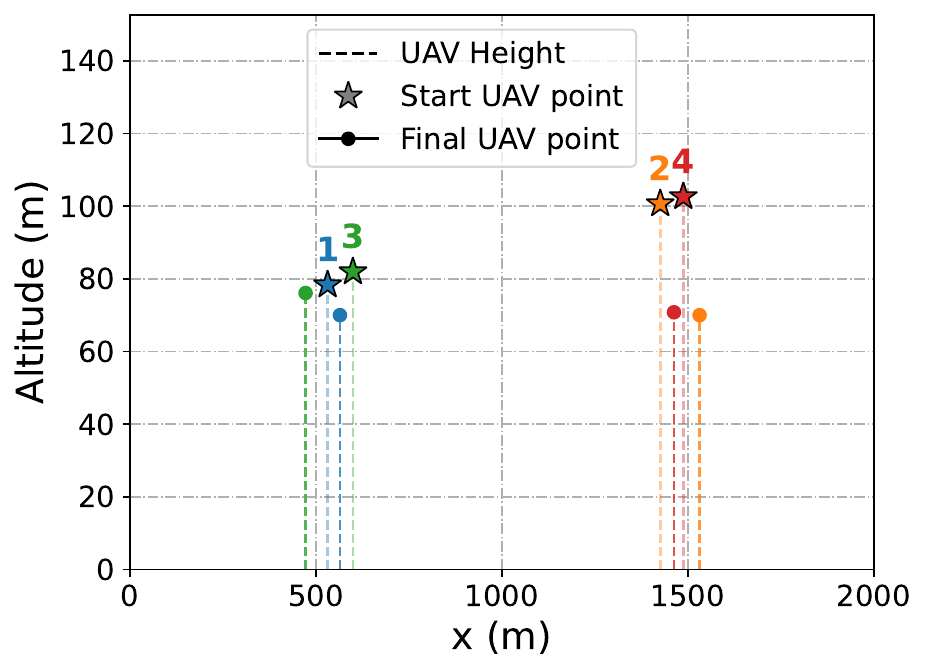} \\
    \small (a) & 
    \small (b) & 
    \small (c) 
\end{tabular}
\vspace{-5pt}
\caption{(a) 2D configuration with UAVs trajectories, (b) Final UAVs 3D placement, and (c) Final heights of UAVs.}
\vspace{-16pt}
\label{fig_3d_placement_plots}
\end{figure*}

\section{Simulation Results} \label{section4}
To evaluate the performance of our approach, we consider a square target area $\mathcal{A}_c$ of size $2\text{ km} \times 2\text{ km}$ with 80 LoRa end devices organized into four spatial clusters around POIs. The device locations follow a Gaussian distribution centered at each POI with standard deviation of $\sigma_u = 150$ m. The four POIs (cluster centroids) are fixed at $\boldsymbol{\mu}_1=(500,500)\,\text{m}$, $\boldsymbol{\mu}_2=(1500,500)\,\text{m}$, $\boldsymbol{\mu}_3=(500,1500)\,\text{m}$, and $\boldsymbol{\mu}_4=(1500,1500)\,\text{m}$. At the start of each training episode, four UAVs are randomly positioned within their respective clusters. The simulation parameters are summarized in Table~\ref{tab:my-table-parameter}.

We evaluate our proposed MAPPO-HeteroNet against three groups of methods: (i) heterogeneous MARL state-of-the-art, (ii) ablations of our own design, and (iii) non-DRL baselines. First, for the MARL state-of-the-art, we include HAPPO~\cite{yifan_JMLR_HARL}, a heterogeneous-agent actor-critic with sequential policy updates and trust-region guarantees for monotonic joint improvement, and HAA2C~\cite{yifan_JMLR_HARL}, a heterogeneous asynchronous advantage actor-critic with type-specific networks and asynchronous updates for stability. Second, the ablation scenarios isolate the effect of vertical-layer assignment while keeping the device count and spatial distribution fixed. Specifically, MAPPO-UG-Net places all 80 end devices in the underground layer, whereas MAPPO-G-Net deploys all devices to the ground layer. These are compared against the proposed MAPPO-Hetero-Net configuration, which includes a total of 80 end devices equally divided into 40 underground and 40 ground LoRa end devices. Finally, we consider two non-DRL baselines. i.e., a Random approach, which samples UAV actions uniformly at random (movement and PHY parameters), providing a lower bound, and a Fixed+Heuristic approach, where UAVs remain fixed at their POI centers while spreading factor and transmission power parameters are assigned using distance-based allocation.
\begin{table}[tb]
    \centering
    \caption{Simulation setup.}
    \vspace{-0.5em}
    \label{tab:my-table-parameter}
    \begin{adjustbox}{max width=0.98\columnwidth}
    \begin{tabular}{c|c|c|c}
        \toprule
        \textbf{Sym.} & \textbf{Value} & \textbf{Sym.} & \textbf{Value} \\
        \midrule
        $V$   & $80$  &$U$ & $4$   \\
        $[h_{\min}, h_{\max}]$ & $[70,150]$ m   & $h_v^{0}$, $h_v^{1}$ & $-0.4$ m, $0$ m \\
        $f$ & $868$ MHz & $BW_v$ & $125$ kHz  \\
        $\phi$, $\varphi$ & $4.88$, $0.43$& $\eta^\mathrm{G2A}_{\mathrm{LoS}}$, $\eta^\mathrm{G2A}_{\mathrm{NLoS}}$ & $0.1$ dB, $21$ dB  \\
        $c_0$ & $3 \times 10^8$ m/s & $\sigma^{2}$, $\sigma_u$  &  $-120$ dBm, $150$ m \\
        $\mathbf{\Phi}$ & $\{7, 8, 9, 10, 11, 12\}$ & $\mathbf{P}$ & $\{2, 5, 8, 11, 14\}$ dBm  \\
        $k_{\text{ind}}$, $W_u$ & $0.11$, $20.0$ N & $\bar{j}$, $\rho$, $A_{\text{rotor}}$ & $4$, $1.168$ kg/m$^3$, $0.214$ m$^2$  \\
        $\epsilon_0$ & $8.854\times10^{-12}$ F/m  & $\eta^0$ & $2.0$ \\
        $\epsilon'$, $\epsilon''$ & $18.2030$, $0.16287$ & $m_v$, $d_p$ & $0.20$, $0.41146$ m  \\
        $\mu_0$, $\mu_r$ & $4\pi\times10^{-7}$ H/m, $1.0$ & Clay \& Sand & $10\%$ \& $90\%$ \\
        $\alpha_{\text{actor}}$, $\alpha_{\text{critic}}$ & $3\times10^{-4}$, $5\times10^{-4}$  & $\epsilon_{\text{clip}}$, $\gamma$, $T$, $T_{\text{tot}}$ & $0.2$, $0.95$, $50$, $3M$ \\
        $\beta_{\text{ent}}$, $\omega$ & $0.01$, $0.3$ & Max steps/episode & $100$ \\
        Seed & $\{0, 44, 182, 235\}$  & Architecture  & MLP (2 layers, $[128, 128]$)  \\
        Optimizer & Adam  & Activation & ReLU \\
        \bottomrule
    \end{tabular}
    \end{adjustbox}
    \vspace*{-2.5em}
\end{table}

\begin{figure*}
\centering
\begin{tabular}{ccc}
    \includegraphics[width=0.315\textwidth]{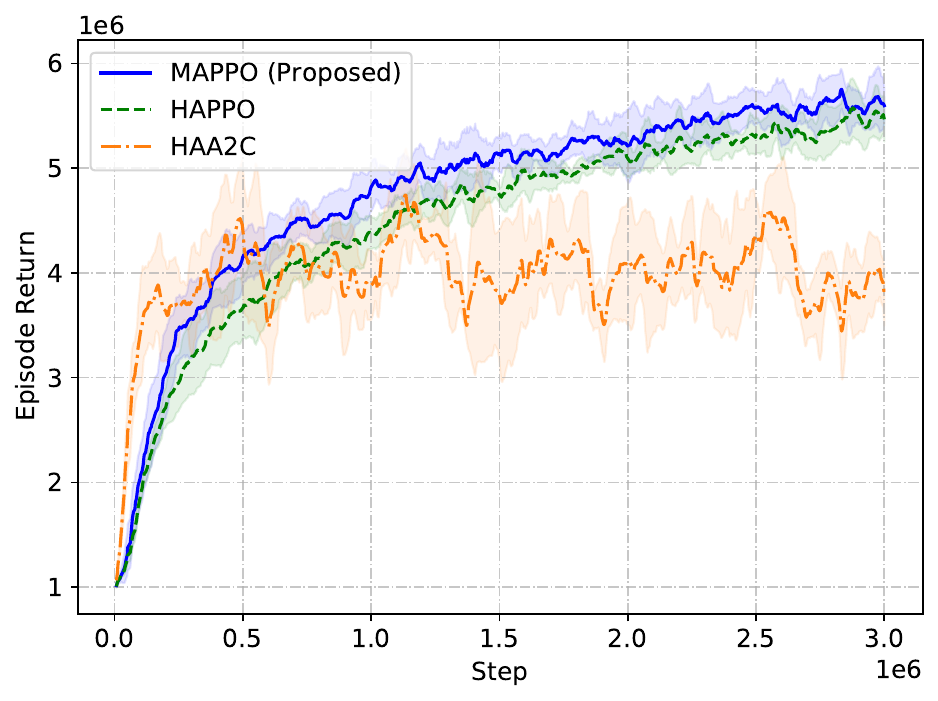} & 
    \includegraphics[width=0.315\textwidth]{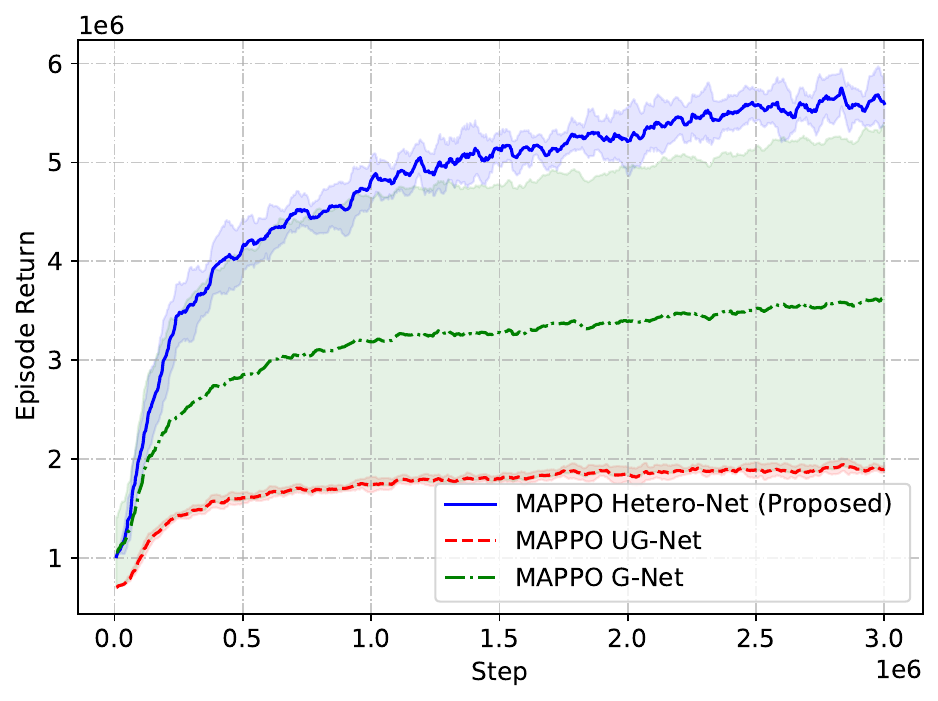} & 
    \includegraphics[width=0.315\textwidth]{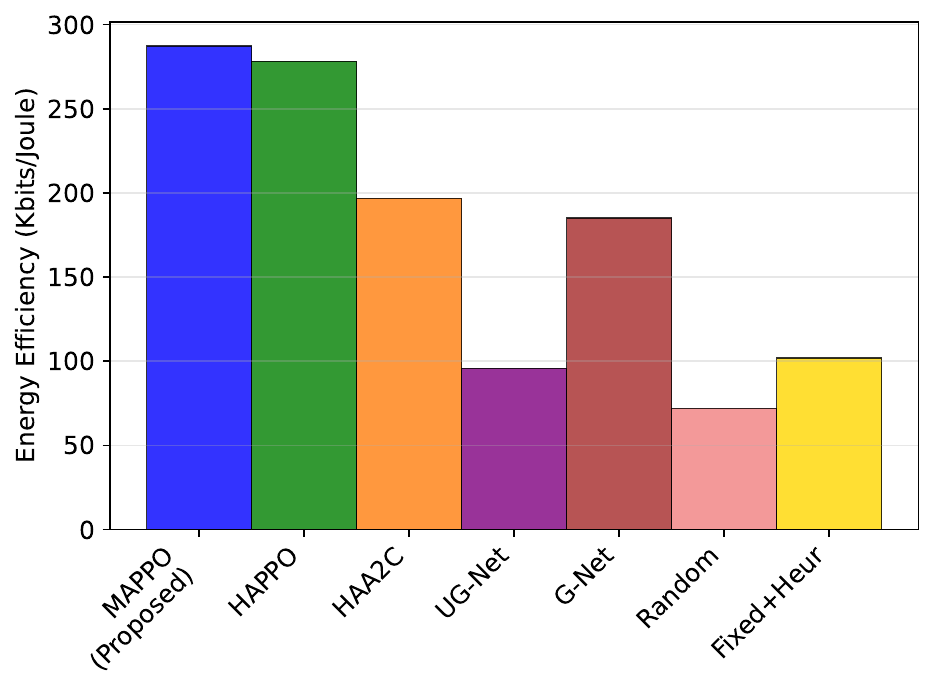} \\
    \small (a) & 
    \small (b) & 
    \small (c) 
\end{tabular}
\vspace{-5pt}
\caption{(a) Training rewards over environment steps. (b) Ablation study. (c) Energy efficiency with benchmarking schemes.}
\vspace{-15pt}
\label{fig_Training_rewards_ee}
\end{figure*}

Fig. \ref{fig_3d_placement_plots} illustrates the initial and final configurations of the four flying gateways in the proposed Hetero-Net scenario. Specifically, Fig. \ref{fig_3d_placement_plots}(a) shows each UAV beginning from a random location within its cluster and moving towards the barycenter of its assigned LoRa end devices. As depicted, dashed lines indicate the associations between LoRa end devices and their serving flying gateways. Furthermore, Fig. \ref{fig_3d_placement_plots}(b) presents the final 3D placement. Here, the flying gateways settle near the centers of their respective end-device clusters. They position themselves to serve both ground WSN LoRa end devices and underground WUSN LoRa end devices, while also improving line-of-sight conditions for the ground nodes. Fig. \ref{fig_3d_placement_plots}(c) depicts the starting point, final point, and flying height of our deployed flying gateways in a 2D plane. As shown from the figure, the UAVs adopt different heights to improve coverage and reduce interference.
 
Fig.~\ref{fig_Training_rewards_ee}(a) presents the convergence of cumulative training rewards for the DRL-based algorithms. As observed from the figure, the proposed MAPPO-Hetero-Net achieves the highest episode return, followed by the HAPPO algorithm. 
In contrast, HAA2C initially starts with higher rewards but begins to fluctuate around $0.5 \times 10^6$ steps and shows greater variability. This behavior is likely due to its asynchronous updates and less effective handling of the dynamic characteristics of the environment. In Fig.~\ref{fig_Training_rewards_ee}(b), the ablation study results are shown, comparing the heterogeneous setup against homogeneous configurations. Specifically, the performance of MAPPO-UG-Net (underground-only) and MAPPO-G-Net (ground-only) drops significantly, highlighting the importance of vertical-layer diversity in improving cooperative learning and network adaptability. Finally, Fig.~\ref{fig_Training_rewards_ee}(c) compares the system energy efficiency across all benchmark schemes. As can be seen from the figure, the proposed MAPPO-Hetero-Net achieves the highest energy efficiency, outperforming G-Net and UG-Net by 55.81\% and 198.49\%, respectively, and surpassing traditional non-DRL baselines such as Random and Fixed+Heuristic by 298.73\% and 181.35\%, respectively.
\vspace{-3pt}

\section{Conclusion} \label{section5}

In this paper, we investigated uplink data collection in heterogeneous LoRa networks integrating ground-based WSNs and underground WUSNs. We proposed a Hetero-Net framework that jointly optimizes the spreading factor, transmission power, and 3D placement of UAV-mounted gateways to maximize system energy efficiency. The problem was formulated as a POSG and addressed using a MAPPO-based MARL approach within the CTDE paradigm. The proposed model incorporates quality-of-service constraints and the reliability of both G2A and UG2A channel models. Simulation results show that the proposed Hetero-Net framework significantly improves energy efficiency and learning performance compared to homogeneous network designs and non-DRL baselines. Future work will focus on experimental validation in more complex environments, as well as extending the framework to include ground-vehicle-mounted gateways and multi-layered Non-Terrestrial Network (NTN) architectures.

\balance
\bibliographystyle{IEEEtran} 
\bibliography{ref}
\end{document}